\documentclass[amssymb,prl,aps,twocolumn,amsmath,showpacs]{revtex4} 
\usepackage[T1]{fontenc}
\usepackage{graphicx}
\usepackage{color}
\usepackage{amsmath}
\usepackage{ulem}
\usepackage{hyperref}

\newcommand{\Skt}{S_k^T}
\newcommand{\Skq}{S_k^Q}

\newcommand{\Lobs}{L_{\rm{obs}}}
\newcommand{\Lpix}{L_{\rm{pix}}}
\newcommand{\npix}{n_{\rm{pix}}}

\newcommand{\gone}{g^{(1)}} 
\newcommand{\gtwo}{g^{(2)}}

\newcommand{\Var} {\langle \delta N^2 \rangle}
\newcommand{\Varmtot}{ {\langle \delta N^2 \rangle}_m^{\rm{tot}}}

\newcommand{\N} {\langle N \rangle}

\newcommand{\varnk} { \langle \delta n_k^2  \rangle}

\newcommand{\Uk}{\mathcal{U}_k}

\newcommand{\ldb}  {\lambda_{\rm dB}}
\newcommand{\om} {\omega_{\perp}}

\newcommand{\ompi} {\omega_{\perp}/2\pi}
\newcommand{\omlpi} {\omega_{\parallel}/2\pi}

\newcommand{\Rb} {^{87}\rm{Rb}}

\newlength{\halfcolumnwidth}
\begin{document}

\title{Direct observation of quantum phonon fluctuations in a one dimensional Bose gas}
\date{\today}
\author{Julien Armijo}
\email{julienarmijo@gmail.com}
\affiliation{Laboratoire Charles Fabry, Institut d'Optique, UMR8501 du CNRS, 91127 Palaiseau Cedex, France}

\begin{abstract}

We report the first direct observation of collective quantum fluctuations in a continuous field.
Shot-to-shot atom number fluctuations in small sub-volumes of a weakly interacting ultracold atomic 1D cloud are studied using \textit{in situ} absorption imaging and statistical analysis of the density profiles. 
In the cloud centers, well in the \textit{quantum quasicondensate} regime, the ratio of chemical potential to thermal energy is $\mu/ k_B T\simeq4$, and, owing to high resolution, up to 20\% of the microscopically observed fluctuations are quantum phonons. Within a non-local analysis at variable observation length, we observe a clear deviation from a classical field prediction, which reveals the emergence of dominant quantum fluctuations at short length scales, as the thermodynamic limit breaks down.

\end{abstract}

\maketitle

At temperature $T=0$, classical thermodynamics predicts the complete absence of excitations, however, due to the Heisenberg uncertainty principle, quantum observables never fully come to rest.
Quantum, i.e. vacuum fluctuations are key to the understanding of quantum electrodynamics effects as fundamental as spontaneous emission \cite{CohenPIPA}, the Lamb shift \cite{Lamb47}, or Casimir-Polder forces near surfaces \cite{Casimir45, Casimir48, Sukenik93}, but also, Hawking radiation near black holes \cite{Hawking75}, quantum phase transitions \cite{Sachdev00}, etc.
In matter fields, quantum fluctuations govern the correlations properties at low temperature. They cause quantum depletion in Bose-Einstein condensates, bringing corrections to their equation of state (EoS) \cite{Lee57, Navon11}, and are dramatically enhanced in reduced dimensions \cite{Bloch08}. In 1D systems, they destroy long range order and prevent Bose-Einstein condensation even at $T=0$ \cite{Petrov04}.

So far, quantum fluctuations in continuous fields have been detected only indirectly, from their \textit{macroscopic} consequences at the thermodynamic scale \cite{Sukenik93, Navon11}, while their microscopic observation has remained elusive \cite{Hofferberth08, Stimming10}.
Only recently, density imaging of ultracold atomic clouds has allowed to \textit{microscopically} image a quantum fluctuating field (more precisely, its modulus) in discrete lattice systems \cite{Endres11}. In this Letter, we report microscopic observation of quantum fluctuations for the first time in a \textit{continuous} field.

Statistical analysis of in-situ density fluctuations has indeed become a prominent tool of investigation. In dimension $D$, the variance of atom number $\Var$ 
in a sufficiently large volume $\Delta^D$ is the same as for an infinite system in thermodynamic equilibrium at the local density $n$ and chemical potential $\mu$ \cite{Armijo10, Klawunn11} :
\begin{equation}
\Var=  \Delta^D k_B T (\partial n/\partial \mu)_T,
\label{eq.fluctudiss}
\end{equation}
where $k_B$ is the Boltzmann constant, and $(\partial n/\partial \mu)_T/n^2$ the isothermal compressibility, derived from the EoS $n(\mu, T)$.
This thermodynamic regime has allowed to observe bosonic bunching \cite{Esteve06, Armijo10} and fermionic antibunching \cite{Sanner10, Muller10} in ideal gases. In weakly repulsive Bose gases, the suppression of density fluctuations, which defines the quasicondensate regime, was detected directly in 1D \cite{Esteve06} and 2D \cite{Hung11}. Recent thermodynamic studies addressed the phase diagram for quasicondensation in 1D and elongated 3D Bose gases \cite{Armijo11}, the universality of the 2D Bose gas \cite{Hung11}, and, in optical lattices, the superfluid to Mott insulator transition \cite{Gemelke09}.
In 1D Bose gases at $k_B T < \mu$, subpoissonian density fluctuations, i.e., antibunching \cite{Jacqmin11} provided thermodynamic evidence of the \textit{quantum quasi-condensate} regime \cite{Kheruntsyan03, Bouchoule11}.

Thermodynamic measurements along Eq.~\ref{eq.fluctudiss} require that the relevant excitations under observation have occupation numbers $n_k \gg 1$, so that thermal fluctuations dominate, and fluctuations are \textit{classical}, i.e. proportional to $T$, as in Eq.~\ref{eq.fluctudiss}, with no quantum contribution. This is valid only in the thermodynamic limit
\begin{equation}
\Lobs \gg  l_c^T,
\label{eq.thermocond}
\end{equation}
where $l_c^T$ is the thermal correlation length of density fluctuations, and $\Lobs$ is the length scale at which the system is probed.
If $\Lobs < l_c^T$, measured fluctuations can deviate from Eq.\ref{eq.fluctudiss} \cite{Hung11, Klawunn11}, and, if $n_k<1$, quantum fluctuations dominate.
In this Letter, we analyze density fluctuations in 1D Bose gases with $\mu/k_B T\simeq 4$, reaching the crossover regime $l_c^T \sim \Lobs$. 
Here, quantum fluctuations have a \textit{sizeable} contribution. Still, quantum and thermal non-thermodynamic effects cancel almost exactly, and Eq.~\ref{eq.fluctudiss} apparently holds.
However, quantum fluctuations are non-extensive \cite{Klawunn11, Jacqmin11}, scaling only \textit{logarithmically} with $\Lobs$, while thermal fluctuations scale \textit{linearly} (as in Eq.~\ref{eq.fluctudiss}, with $\Lobs \simeq \Delta$). 
Hence, truly thermodynamic, i.e., purely thermal fluctuations, are observed only for $\Lobs \gg l_c^T$.
By varying $\Lobs$, we monitor the breakdown of the thermodynamic limit at small $\Lobs$ as a clear deviation from a classical field theory that ignores quantum fluctuations. In the line of \cite{Stimming10}, this criterium proves our microscopic observation of quantum fluctuations.

Our experiment uses $\Rb$ atoms in micro-magnetic traps on an atom chip. The transverse and longitudinal trapping frequencies are $\ompi= 3.3$ kHz and $\omlpi = 5.5$ Hz respectively. After forced rf evaporative cooling and thermalization for 800 ms, an absorption picture is recorded on a CCD camera.
After hundreds of realizations, fluctuations in the density profiles are analyzed as detailed in \cite{Armijo10}. 
For each profile and pixel of length $\Delta = 4.5\ \mu$m, we extract the atom number fluctuation $\delta N = N - \N$ where $\N = n \Delta$ is the mean atom number.
\textit{True} atom number variances $\Var$ are inferred from \textit{measured} variances $\Var_m$ using the thermodynamic relation $\Var_m = \kappa_2 \Var$, where $\kappa_2$ is a reduction factor due to the finite $rms$ imaging resolution $\delta$.
Assuming a gaussian imaging response, $\delta$ is determined precisely from the correlations measured between neighboring pixels \cite{Armijo10, supMat}.

\begin{figure}
\begin{center}
\includegraphics[width=8.6cm]{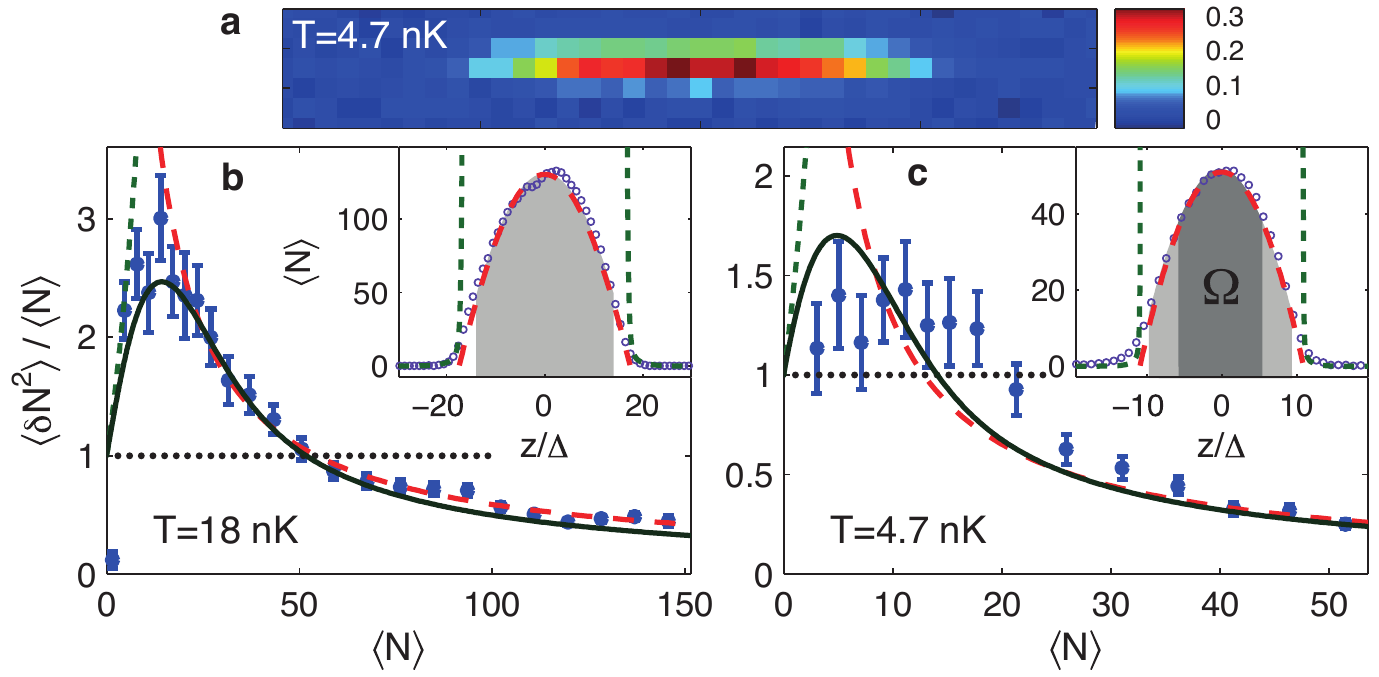}
\caption{(Color online). Microscopic fluctuations and thermodynamic theories, at $T=18$ nK (b) and $T=4.7$ nK (a, c).
(a) : Typical absorption picture (optical density). (b, c) Normalized fluctuations and average profiles (insets).
Theoretical curves : classical shot noise (dotted), predictions from Eq.~\ref{eq.fluctudiss} and the ideal Bose gas (short-dashed), quasi-condensate (long-dashed), and Yang-Yang EoS (solid). 
Grey areas indicate in the profiles indicate the quantum quasi-condensate regime, the dark-grey area in the $T=4.7$ nK profile is the zone $\Omega$ where the analysis of Fig.~4 is carried.}
\label{fig.antibunch}
\end{center}
\end{figure}

To introduce our data, we show in Fig.~1 a typical picture (a), average profiles, and relative fluctuations, for clouds deep in the 1D regime, with $k_B T/\hbar \om = 0.11$ (b) and 0.03 (c).
As in \cite{Armijo10, Armijo11}, $T$ is measured accurately by fitting the fluctuations in the cloud centers with Eq.~\ref{eq.fluctudiss} and the Gross-Pitaevski (G-P) quasicondensate EoS in the 1D-3D crossover $\mu = \hbar \om (\sqrt{1 + 4 na} -1)$  \cite{Mateo08} (dashed curve, see also Fig.~3), where $a=5.3$ nm is the 3D scattering length.
At low densities, fluctuations are superpoissonian, i.e., exceed the shot noise (dotted line). They reach a maximum at the transition to the quasicondensate regime, around the density $n_{\rm{qc}}= [m(k_BT)^2)/\hbar^2g]^{1/3}$ \cite{Armijo11}, in good agreement with the exact Yang-Yang solution for the 1D Bose gas \cite{YangYang69} (solid curve)
\footnote{In the cloud wings  of Fig.~1c ($\N<30$), slight inaccuracies could stem from important density gradients that make the local density approximation questionable \cite{supMat}.}.
At higher densities, for $\mu > k_B T$ (light grey areas in profiles), quantum fluctuations start dominating \textit{thermodynamically} and fluctuations turn subpoissonian, as explained in \cite{Jacqmin11}.
In this Letter, the important region is the cloud center at $T=4.7$ nK (zone $\Omega$), where $\Var/\N \simeq k_B T/\mu$ reaches 0.26, and where we detect quantum phonons \textit{microscopically}, as explained below.

To analyze fluctuations in the cloud centers, we use Bogoliubov theory, valid for weakly interacting quasicondensates \cite{Mora03}, and the 1D G-P EoS $\mu=gn$, where $g=2 \hbar \om a$ is the 1D coupling constant. 
This is appropriate since the dimensional correction to the G-P prediction is only 10\% in zone $\Omega$, and the 1D interaction parameter is $\gamma = g m/\hbar^2 n \lesssim 0.05$.
Bogoliubov excitations have energies $\epsilon_k = \hbar^2 k^2/2m\sqrt{1+4/k^2\xi^2}$, where $\xi=\hbar / \sqrt{m \mu}$ is the healing length, and thermal occupation numbers $n_k=1/(e^{\epsilon_k /k_B T}-1)$. 
Noting $f_k =1/\sqrt{1+4/k^2\xi^2}$, the spectrum of density fluctuations is 
\begin{equation}
 \varnk =n S_k =n (S_k^Q + S_k^T)  ,
\label{eq.varnk}
\end{equation} 
where $S_k^Q= f_k$ and $S_k^T=2f_k n_k$ are the quantum and thermal static structure factors, both plotted in Fig.~2c. 
Note that, in $S_k=2f_k (n_k +1/2)$, $S_k^Q$ is the exact analog of the zero-point energy term in the harmonic oscillator spectrum.
We also show in Fig.~2d the second order correlation function $\gtwo(z) =1+ \int \frac{dk}{2\pi} e^{i k z} (S_k -1)$.

\begin{figure}
\begin{center}
\includegraphics[width=8.6cm]{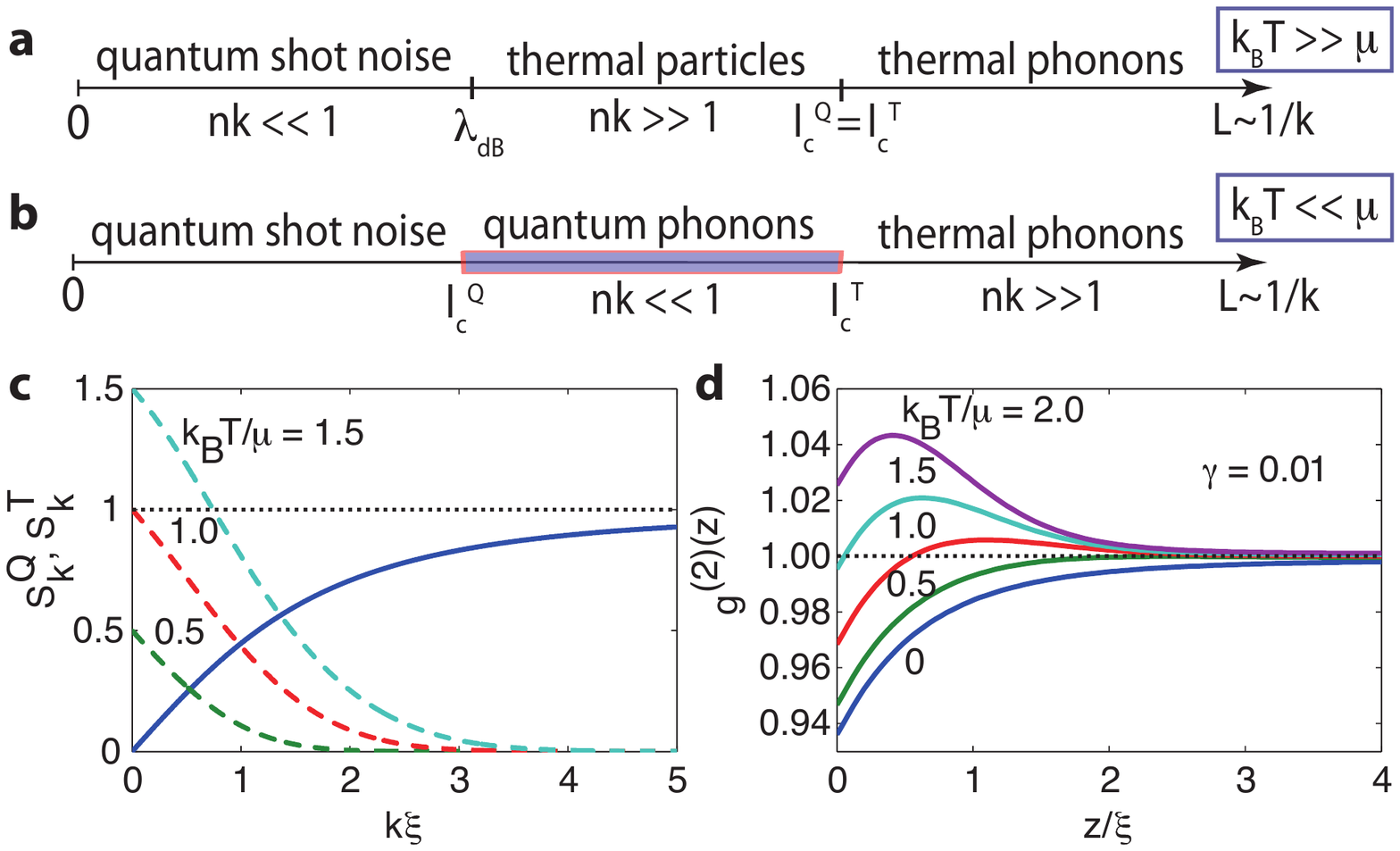}
\caption{(Color online). Quantum \textit{vs} thermal fluctuations in 1D quasicondensates. (a, b) Fluctuation regimes for thermal (a) and quantum (b) quasicondensates. (c) $S_k^Q$ (solid) and $S_k^T$ (dashed) in Bogoliubov theory for various $k_B T /\mu$. 
(d) $\gtwo(z)$ for $\gamma=0.01$.}
\label{fig.bogo}
\end{center}
\end{figure}

Three competing terms determine the fluctuation regimes in quasicondensates.
For $k \xi \gg 1$, $S_k^Q\simeq 1$ is the trivial autocorrelation "shot noise" \cite{Esteve06, Armijo10} of uncorrelated particles, for which $\gtwo(z) \equiv1$. At smaller $k$, repulsive interactions give a negative contribution to $S_k-1$ and to $\gtwo(z)-1$ \cite{Jacqmin11}. Finally, in $S_k^T$, statistical bosonic bunching brings a positive contribution.
Let us now focus on the quantum and thermal density correlation lengths $l_c^Q$ and $l_c^T$, respectively defined as the inverse widths of $S_k^Q$ and $S_k^T$.
For quantum fluctuations, equally present in all modes $k$, one has $l_c^Q = \xi$ \cite{Deuar09}. For thermal fluctuations, two cases need be distinguished. 

In \textit{thermal quasi-condensates} ($k_B T \gg \mu$, see Fig.~2a), $S_k^T$ decays fast for $k l_c^Q \gg1$, so one has a single $l_c=l_c^T=l_c^Q$ \cite{Bouchoule11, Deuar09}. However thermal fluctuations dominate, i.e. $n_k \gg 1$, for length scales $L\sim 1/k \gg \ldb$, where $\ldb=\sqrt{2 \pi \hbar^2/mk_B T}$ is the thermal de Broglie wavelength.
Three regimes are thus present \cite{Klawunn11}.
For $L \gg l_c$, thermal phonons dominate : this is the thermodynamic regime, with superpoissonian fluctuations.
For $\ldb \ll L \ll l_c$, thermal particles dominate, but correlations are partly lost
\footnote{This regime explains the recent observation of fluctuations well below the prediction of Eq.~\ref{eq.fluctudiss} in thermal 2D quasicondensates, with $\Lobs \lesssim l_c^Q$ \cite{Hung11, Hung112, Klawunn11}.}. 
Finally, for $L \ll \ldb$, only the poissonian quantum shot noise is detected.

In \textit{quantum quasicondensates} ($k_B T \ll \mu$, see Fig.~2b), noting $c=\sqrt{\mu/m}$ the speed of sound, $n_k$ and $S_k^T$ are both suppressed at length scales lower than
\footnote{One obtains $l_c^T$ requiring that the phonon energy $\hbar c k \gg k_B T$ (and thus $n_k \ll1$) for wavevectors $k \gg1/l_c^T$.}
\begin{equation}
l_c^T  =\frac{\hbar c}{k_B T} =  l_c^Q \frac{\mu}{k_B T} ,
\label{eq.lQ}
\end{equation}
which, to our knowledge, has been proposed only recently \cite{Gangardt10, Stimming10, Klawunn11}. 
The familiar regimes of poissonian shot noise and thermodynamic thermal phonon fluctuations (now subpoissonian) are still present. However, in the range $l_c^Q \ll L \ll l_c^T$, one now has a crucial new regime (highlighted in Fig.~2b), where  $n_k \ll1$ and \textit{quantum phonons} govern the physics.

To compare Bogoliubov theory to our data, we use the imaging transfer function $\Uk=2 \frac{(1-\cos(k\Delta))}{ k^2} e^{-k^2\delta^2}$ \cite{Jacqmin11}, obtained from our gaussian optical response model \cite{Armijo10, supMat}, and we compute, for $j=Q,T$,
\begin{equation}
\Var_m^{j} = \int \frac{dk}{2\pi} n S_k^{j} \Uk.
\label{eq.varFourier}
\end{equation}
$\Uk$ is peaked at $k=0$, and its inverse width is $\Lobs \sim \rm{max} \{ \delta, \Delta\}$. 
The thermodynamic limit Eq.~\ref{eq.thermocond} is thus equivalent to measuring $\Var_m = \kappa_2 \N S_0$ \cite{Klawunn11, supMat}, i.e., to only probe the contribution $S_0=S_0^T$ of thermal phonons, always proportional to $T$ (see Fig.~2c). In other words, since $S_0=1+\int dz(\gtwo(z)-1)$, a thermodynamic observation probes only the \textit{integral} of $\gtwo(z)-1$, without resolving its microscopic details (see Fig.~2d).

\begin{figure}
\begin{center}
\includegraphics[width=8.6cm]{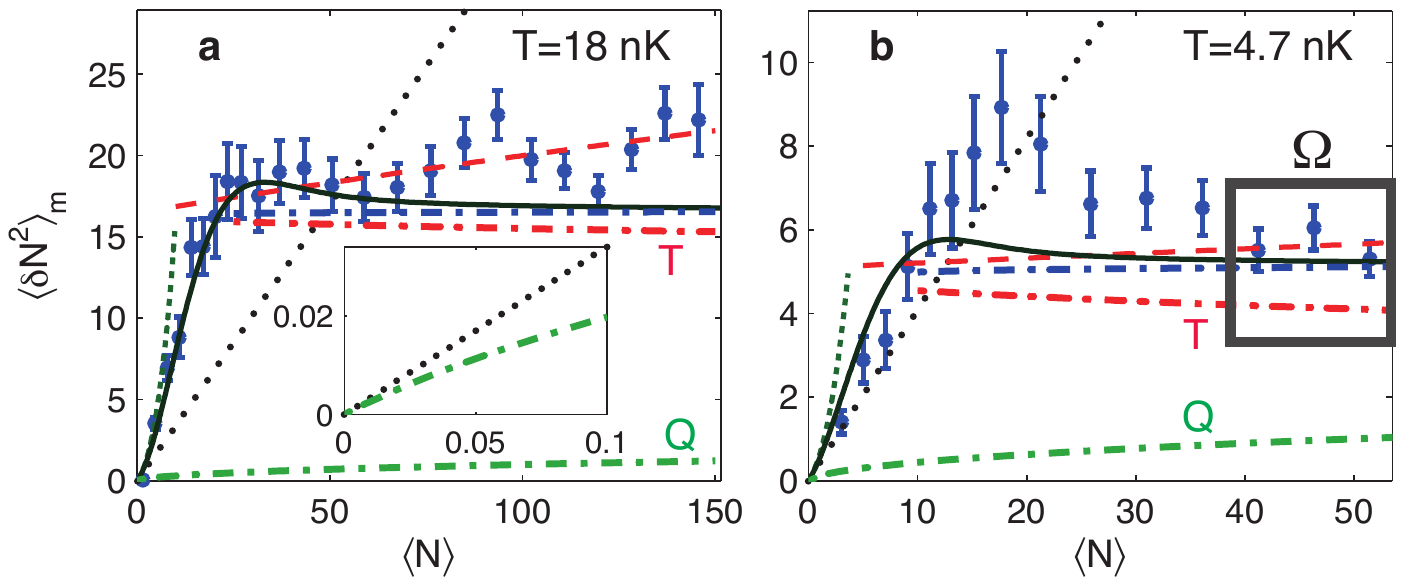}
\caption{(Color online). Quantum \textit{vs} thermal contributions in the measured fluctuations $\Var_m$ (same data as in Fig.~1).
Dot-dashed lines are the spectral 1D Bogoliubov predictions $\Var_m^Q$ (low-lying green), $\Var_m^T$ (middle red) and $\Varmtot$ (upper blue). 
Thermodynamic predictions as in Fig.~1 are here rescaled by $\kappa_2=0.34$ (a) and 0.41 (b) for $\delta=3.5$ and $2.8\ \mu$m, respectively.
The square in (b) is the zone $\Omega$.}
\label{fig.varbogo}
\end{center}
\end{figure}

In Fig.~3, we compare the measured fluctuations $\Var_m$ to the  spectral 1D Bogoliubov predictions $\Var_m^Q$ and $\Var_m^T$, and their sum $\Varmtot$ (dot-dashed lines). 
At low densities, one has $ l_c^Q \gg \Lobs$, which explains that $\Var_m^Q$ follows the shot linear noise prediction (see inset to Fig.~3a).
At higher densities,  $\Var_m^Q$ is \textit{reduced} by repulsive interactions
\footnote{On the contrary, quantum \textit{phase} fluctuations are \textit{increased} by interactions, causing stronger condensate depletion \cite{Petrov04}.}.
At $T=18$ nK, the ratio $\Var_m^Q/\Varmtot$ is $7\%$ and cannot be resolved, as in \cite{Jacqmin11}. However, at $T=4.7$ nK, it reaches 20\% in the cloud center, which now exceeds the experimental uncertainty.
Thus, the contribution of quantum fluctuations is here \textit{sizeable}, i.e. non-negligible, in each pixel, in contrast to the thermodynamic regime.
Yet, $\Varmtot$ is still in good agreement with the 1D thermodynamic Yang-Yang prediction (solid line) 
\footnote{This is also true for the 1D G-P prediction which is $3\%$ below the exact Yang-Yang prediction. The small correction (at lowest order, $\sqrt{\gamma}/2$) stems from quantum fluctuations and is the 1D equivalent of the Lee-Huang-Yang correction in 3D \cite{Lee57, Mora03}}.
This is because $\Skq$ and $\Skt$ have opposite slopes $\pm k\xi/2$ at small $k \xi$ for all $T$ (see Fig.~2.c), so that non-thermodynamic (i.e., finite $k$ \cite{Klawunn11}) quantum and thermal contributions cancel each other at first order.

To confirm our detection of quantum fluctuations, we turn to a \textit{non-local} analysis. 
Since quantum fluctuations scale only \textit{logarithmically} with $\Lobs$ \cite{Jacqmin11, Klawunn11}, thermal fluctuations always dominate for $\Lobs \gg l_c^T$, and fluctuations are well described by a classical field model (CFM) that ignores the quantum term $S_k^Q$ \cite{Stimming10},  On the other hand, a CFM is expected to fail for $\Lobs \lesssim l_c^T$.
To check this, we focus on the zone $\Omega$, i.e., the 3 central bins $39<\N<54$ at $T=4.7$ nK, containing 58\% of the atoms (see inset to Fig.~1d and Fig.~3b).
There, the profile is the flattest, many data points are available, and error bars are the smallest. The cloud center is indeed the most reliable fraction of the data.

\begin{figure}
\begin{center}
\includegraphics[width=8.6cm]{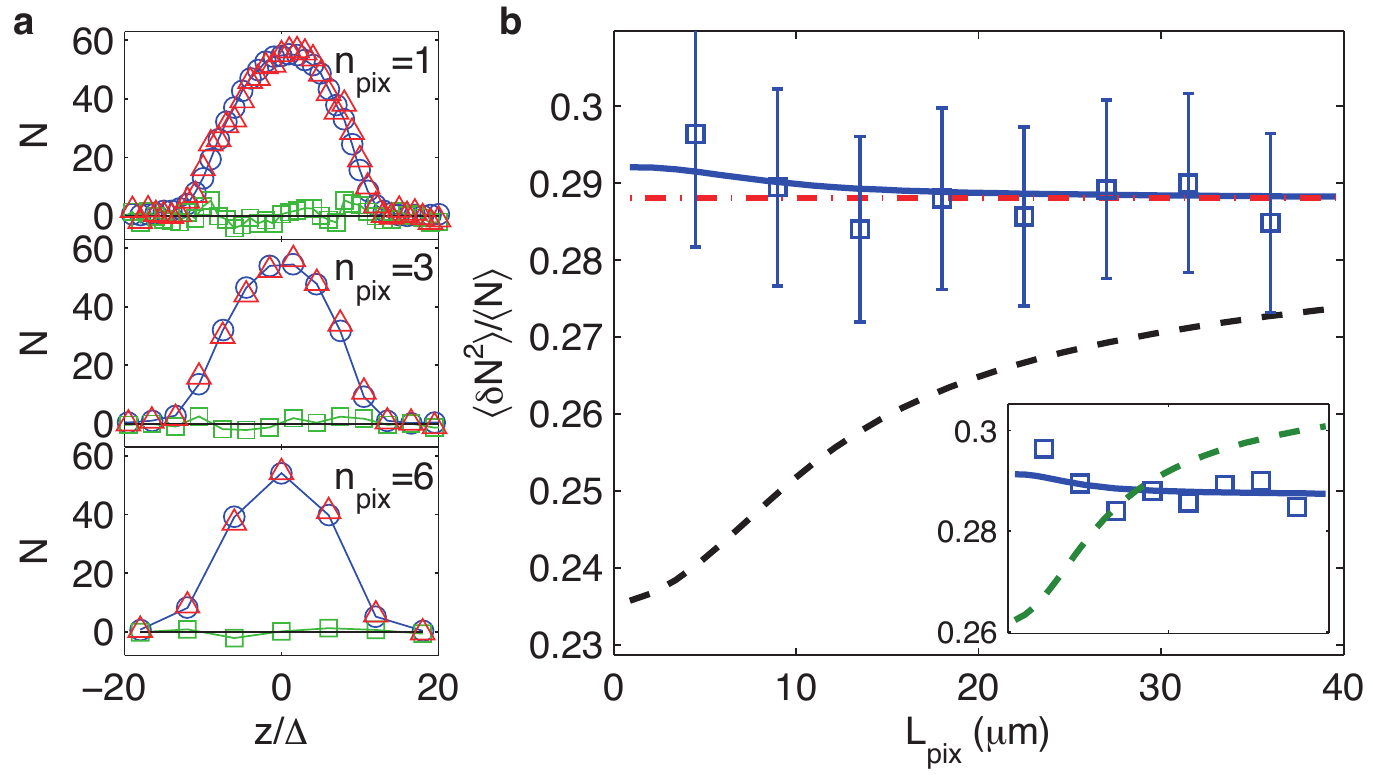}
\caption{(Color online). Fluctuations at variable length scales in the cloud center (zone $\Omega$) at $T=4.7$ nK.
(a) Typical profile $N$ (triangles), reference profile $\N$ (circles) and fluctuations $\delta N$ (squares), for three pixel sizes $\Lpix$.
(d) $\Var/\N$  in zone $\Omega$ compared to full Bogoliubov (FBM) prediction $\Var_m^{\rm{tot}}/\kappa_2$ (solid), classical field (CFM) prediction $\Var_m^T/\kappa_2$ (dashed) and thermodynamic G-P prediction (dot-dashed) at $T_{\rm{1D}}^*=5.0$ nK.
Inset : same data fitted with FBM at $T_{1D}^*=5.0$ nK (solid) or CFM at $T_{1D}^C=5.5$ nK (dashed).}
\label{fig.binning}
\end{center}
\end{figure}

To vary $\Lobs$, we merge the imaging pixels in macro-pixels of variable size $\npix= 1$ to 8, and compute fluctuations accordingly (see Fig.~4a and \cite{supMat}).
Figure~4b shows $\Var/\N$ obtained in the zone $\Omega$, for $\Lpix=\Delta \npix= 4.5$ to 36 $\mu$m.
The key criterium, as in \cite{Stimming10}, is to compare the data either to a full Bogoliubov model (FBM) that includes the quantum term (solid line) or to a CFM that ignores it (dashed line). 
At large $\Lpix \ (\simeq \Lobs)$, both models converge as expected towards the thermodynamic 1D G-P prediction (dot-dashed line), with a \textit{logarithmic} vanishing of the quantum contribution.
At short $\Lpix$, the data clearly deviate from the CFM prediction, which displays a noticeable dip.
Note that the gap saturates at small $\Lpix$, due to the finite optical resolution that cuts off high $k$ fluctuations, which confirms that shot noise is irrelevant in our data.

The theoretical predictions in Fig.~4b are computed at $T_{\rm{1D}}^*=5.0$ nK, that we obtain fitting the FBM to the data, with $T$ as only free parameter (see inset to Fig.~4b). 
This fit has an \textit{rms} deviation of only 1.0\%, i.e., much less than the plotted error bars
\footnote{Error bars clearly overestimate the real noise in Fig.~4.b. They give the statistical uncertainty for each point independently, however all are obtained from the same data set, hence most of the noise is correlated and rejected.}
and is thus a very accurate 1D thermometry
\footnote{$T_{\rm{1D}}^*$ slightly exceeds the previous estimate from the 1D-3D G-P EoS, as expected.}.
On the contrary, the fit to the CFM, yielding $T_{\rm{1D}}^{\rm{C}}=5.5$ nK, has a strong systematic error and an \textit{rms} deviation of 5\% (see inset to Fig.~4b).
This clear breakdown of the CFM proves our observation of quantum fluctuations, and reveals the emergence of dominant quantum  phonons at short distances.

As for length scales, in the zone $\Omega$, $l_c^Q = 0.6\ \mu$m and $l_c^T = 2.2\  \mu$m, while $\Lobs$ is determined by $\Delta =4.5\  \mu$m and $\delta = 2.8\ \mu$m. 
Thus, $\Lobs \sim l_c^T$ and this explains our unprecedented observations.
On the contrary, in \cite{Hofferberth08}, at $T=33$ nK, 
one has $l_c^T=0.63\ \mu$m, and $\Lobs \geq \Delta = 10\ \mu$m, i.e., $\Lobs \gg l_c^T$.
All observations in \cite{Hofferberth08} were thus well in the thermal fluctuations regime, indistinguishable from a CFM prediction \cite{Stimming10}.
In the quantum regime $z \ll l_c^T$, the decay of the first order (phase) correlation function $\gone(z)$ is \textit{algebraic}, whereas, in the thermal regime $z \gg l_c^T $, it is \textit{exponential}, over a thermal phase correlation length $l_\phi^T = l_c^T / \sqrt{\gamma} \gg l_c^T$  \cite{Petrov04, Pethick, Armijothesis}. 
The misinterpretation in \cite{Hofferberth08} came from identifying the quantum regime with $z \ll l_\phi^T$.

In summary, we have reported the first microscopic observation of vacuum fluctuations in a continuous field, using a non-local analysis that reveals a clear deviation from a classical field theory. 
Our observation of emerging dominant quantum phonons is a first microscopic insight into the regime of quasi-long range order, i.e, algebraic decay of $\gone(z)$, in the 1D Bose gas \cite{Petrov04}.
We also demonstrated the possibility of imaging vacuum phonon fluctuations in single density pictures, like Fig.~1a.
By further reducing $T$ and $\Lobs$, one could monitor the full crossover from thermal to quantum fluctuations \cite{supMat}.
Dark solitons, which are defects localized over a length scale $\sim l_c^Q$, could also be used as sensitive probes for the microscopic physics of quantum fluctuations \cite{Martin10, Gangardt10}.

 \begin{acknowledgements} We thank I. Bouchoule for discussions and supervision in early stages of the project, K. Kheruntsyan for providing Yang-Yang calculations, 
A. Sinatra for crucial suggestions, Y. Castin for helpful remarks, and F. Werner, T. Yefsah, N. Navon and A. Sinatra for critical comments on the manuscript. 
Work supported by the IFRAF Institute and the ANR grant ANR-08-BLAN-0165-03. 
\end{acknowledgements}

\bibliographystyle{prsty}

\end{document}